\documentclass[a4paper,10pt]{article}
\usepackage[utf8]{inputenc}
\usepackage{amsmath}
\usepackage{mathrsfs}
\providecommand{\del}[2][]{\frac{\delta #1}{\delta #2}}
\providecommand{\sAV}[1]{\langle #1\rangle}
\providecommand{\dd}[2][]{\frac{d #1}{d #2}}
\title{A New Approach to Anomalous Axial Vector Field Theory}
\author{A. K. Kapoor\\ Chennai Mathematical Institute\\
H1 SIPCOT, IT Park, Siruseri\\
Kelambakkam, TN 603103, INDIA}

\begin{document}

\maketitle

\begin{abstract}
In an earlier paper it has been shown that the ultra violet divergence structure of anomalous U(1) axial vector gauge model in the stochastic quantization scheme is different from that in the conventional quantum field theory. Also it has been shown that the model is expected to be renormalizable. Based on the operator formalism of the stochastic  quantization,  a new approach to anomalous U(1) axial vector gauge model is proposed. The operator formalism provides a convenient  framework for analysis of ultra violet divergences, but the  computations in a realistic model become complicated. In this paper a new  approach to do computations in the model is formulated  directly in four dimensions. The suggestions put forward here will lead to simplification in the study of applications of the axial vector gauge theory, as well as those of other similar models.   
\end{abstract}

\vfill
\begin{center}
   ( to appear in Int. J. Mod. Phys. {\bf A} )
\end{center}

\vfill 
\section{Introduction}
In a set of earlier papers a study of anomalous axial vector gauge model coupled to a fermion has been  reported \cite{MPLA19, akk20,akk15}. The main result is that this model 
promises to be renormalizable and has new features which could prove valuable for going beyond the standard model\cite{MPLA19}.  The investigations of models reported so far have been based on the operator formalism of Parisi Wu stochastic quantization method (SQM). The operator formalism entails perturbative computations in a  five dimensional field theory and taking the limit in which the fictitious time, the fifth dimension, goes to infinity, see Sec 2.3. While the operator formalism offers a systematic way of studying ultra violet divergences, actual detailed computations are complex as compared to those in the conventional quantum field theory in four dimensions. 

Following the operator formalism route, in this paper we suggest a scheme that promises to simplify the computations.  
This  proposal has grown out of investigation of the anomalous  axial vector theory using the stochastic quantization scheme proposed by Parisi and Wu \cite{SQM1} which will continue to be the backend of the scheme proposed here. The computation scheme proposed here makes use of the equation obeyed by the steady state Fokker Planck distribution function. In this approach one sets up an equation obeyed  by the generating function directly in four dimensions.    

The plan of this paper is as follows. In the next section we review three available formulations of the  Parisi Wu  stochastic quantization method (SQM) so as to make this article self contained. For more details, the interested reader may see \cite{SQM2,SQM3,SQM4} . In Sec. 3, the results of SQM study of the axial vector gauge theory are summarized. Sec. 4 has the proposal about reduction of equations obeyed by the generating function of the Green functions of the theory in four dimensions. This study concludes  with important remarks in the last section.

\section{Parisi Wu Stochastic Quantization}

We consider a  classical  field theory of a set of  fields 
\(\phi_k(x)\) and described by a classical action 
\(S[\phi_k(x)]\). Here, and everywhere, in this article, \(x\) will collectively denote the four vector \(x^\mu=(x^0, \vec{x})\).
The equations of motion of the classical field are obtained by  principle of stationary action and can be written in a compact form as
\begin{equation}\label{EQ01}
 \del[S]{\phi_k(x)} = 0. 
\end{equation}

The conventional quantum field theory (CQFT) can be summed up by the functional integral representation of the generating function \(Z[J]\) of the Green functions of  QFT:
\begin{equation}\label{EQ02A}
 Z[J] = \int \mathcal{D} \phi e^{i S(\phi_k)} e^{\{i\int dx ~J_k(x) \phi_k(x) \}}.
\end{equation}
Three approaches to the SQM of the field theory described by the action functional \(S[\phi_k]\) will be briefly reviewed. More details can be found in \cite{SQM2,SQM3,SQM4} . The first one is based on Langevin equation. The second approach has the Fokker Planck equation as the starting point. The Fokker Plank approach naturally leads to the third approach to SQM, called the operator approach. The operator approach pioneered by Namiki and Yamanaka to SQM can be viewed as a quantized field theory in five dimensions.  It is this approach that has been found suitable for analysis of the ultra violet divergences. The SQM  is formulated in terms of  the Euclidean action  \(S_E[\phi_k(x)]\) of the model to be studied. The three  formulations of SQM will now be reviewed. 
% 
% \section{Parisi WU Stochastic Quantization}
\subsection{Langevin Equation Formalism}
In the Langevin equation formulation of SQM, the fields 
\(\phi_k(x)\) are viewed as stochastic processes 
\(\phi(x,\tau)\) in fictitious time, also called the fifth time, denoted by 
\(\tau\). The stochastic process is governed by Langevin equation
\begin{equation}\label{EQ03}
 \dd[\phi_k(x,\tau)]{\tau} = -\frac{1}{\gamma_k}\del[S_E]{\phi_k(x,\tau)} + \eta_k(x, \tau).
\end{equation}
The first term in the right side requires some explanation: it is obtained by
computing the functional derivative, \(\del [S_E(\phi_k)]{\phi_k(x)}\) as usual and replacing the field \(\phi_k(x)\) by the stochastic process \(\phi_k(x,\tau)\).
Here \(\eta_k(x,\tau)\) is a Gaussian white noise obeying 
\begin{equation}\label{EQ04} 
 \langle \eta_k(x,\tau)\rangle = 0; \qquad \sAV{\eta_j(x_1,\tau_1) \eta_k(x_{2}, \tau_2)} = 2\gamma_k \delta_{jk} \delta^{(4)}(x_1-x_2), \quad k=1,2...
\end{equation}
In the above equation \(k\) is free index and the last expression does not have a summation over the index \(k\).
The Langevin equation is solved, subject to some chosen initial conditions, and the stochastic processes \(\phi_k(x,\tau)\) is obtained as a functional of the noise \(\eta(x,\tau)\) . Denoting the solution by \(\phi_\eta(x,\tau)\), the equal time correlation functions
(ETCF) obtained by averaging over the Gaussian white noise \(\eta_k\) give the Euclidean Green functions of CQFT. The averaging process can be implemented by  means of functional integral
\begin{eqnarray}\nonumber
\lefteqn{
\sAV{\phi_\eta(x_1,\tau_1), \phi_\eta(x_2,\tau_2),\ldots, \phi_\eta(x_n,\tau_n)}_\eta=}\\ 
&=&\int [\Pi_j\mathcal{D}\eta_j] \big\{\Pi_{k=1}^n\phi_\eta(x_k, \tau_k)\big\} \exp\Big[-\sum_k\Big(\frac{1}{4\gamma_k}\int dx\,d\tau~ \eta_k^2(x,\tau)\Big)\Big].
\end{eqnarray}
where, as already mentioned, \(\phi_\eta(x,\tau)\) is a solution of the Langevin equation over chosen initial conditions.
In the Langevin equation approach, the equal  time averages, \(
\sAV{\phi_\eta(x_1,\tau_1, \phi_\eta(x_2,\tau_2),\ldots, \phi(x_n,\tau_N)}_\eta\), in the limit of 
\(\tau \to \infty\), coincide with the Green functions of the Euclidean theory. If the limit  as  \(\tau\to\infty\) exists, the results will be independent of choice of initial conditions.

\subsection{Fokker Planck Formalism}
In the Fokker Planck equation approach, one aims to introduce a probability distribution \(P[\phi, \tau]\) whose evolution in the fictitious time \(\tau\) approaches \(\exp\big(-S_E[\phi]\big)\) in the limit \(\tau \to \infty\). The Langevin equation \eqref{EQ03} is equivalent to the following Fokker Planck equation
\begin{equation}
 \dd[P{[\phi, \tau]}]{\tau} =  \int dx \sum_k \frac{1}{\gamma_k}\del{\phi_k(x)}\Big[\del{\phi_k(x)}+\del[S_E]{\phi_k(x)} \Big] P[\phi,\tau].
\end{equation}
The Fokker Planck equation can be written as
\begin{equation}\label{EQ77}
 \dd[{P[\phi,\tau]}]{\tau} = - H_{FP} P[\phi,\tau]. 
\end{equation}
where the Fokker Planck Hamiltonian \(H_{FP}\) is given by
\begin{equation}
H_{FP} = \int dx \sum_k \frac{1}{\gamma_k}\del{\phi_k(x)}\Big[\del{\phi_k(x)}+\del[S_E]{\phi_k(x)} \Big] 
\end{equation}

We note that \(\exp \big(-S_E[\phi]\big)\) is a steady state solution of \eqref{EQ77} and, if it is normalizable, it will be an 'eigenfunctional' of the Fokker Planck Hamiltonian with zero as the eigenvalue
\begin{equation}
  H_{FP}\exp\big(-S_E[\phi]\big)=0
\end{equation}
Under the assumptions that zero eigenvalue is non degenerate and discrete, one can show that
\begin{equation}
 \lim _{\tau \to \infty} P[\phi, \tau] =  C \exp\big(-S_E[\phi]\big).
\end{equation}
where \(C\) is a constant. Under the assumptions as indicated above, the equilibrium limit of equal time correlation functions will coincide with
the Green functions of the Euclidean theory.

In the Fokker Planck formalism, the average value of a function \(F(\phi)\) obeys the evolution equation
\begin{equation}
 \dd{\tau} \langle F(\phi, \tau)\rangle_P = \int dx \left( - 
 \del[F]{\phi(x)} 
 \del[S]{\phi(x)} + \frac{\delta^2 F}{\delta \phi(x)^2}\right) \rangle_P.
\end{equation}
Using the Fokker Planck formalism, it is straightforward to show that, in the equilibrium limit, the equal time correlation functions go over to the Green functions of the underlying CQFT, see Sec 2.2 of \cite{SQM2}.

\subsection{Operator Formulation}
A formal similarity of the Fokker Planck equation with imaginary time Schr\"{o}dinger equation cannot be missed. This suggests introduction of a 'stochastic momentum 
operator' \(\pi_k(x)\) conjugate to the field \(\phi_k(x)\) using the identification
\begin{equation}
 \hat{\pi}_k(x) \equiv -\del{\phi(x)}.
\end{equation}
This allows us to cast the Fokker Planck equation in the form
\begin{equation}\label{EQ11}
 \dd[P]{\tau} = - \Big\{\hat{\pi}_k \hat{\pi}_k + \hat{\pi}_k(x) \Big(\del[S_E]{\phi_k}\Big) \Big\} P
\end{equation}
In terms of the stochastic momentum, the Fokker Planck Hamiltonian takes the form
\begin{equation}
   \dd[P]{\tau} = - \mathcal{H}\,\, P[\phi, \tau]
\end{equation}
where \(\mathcal{H}\) is given by
\begin{equation}
 \mathcal{H} = \hat{\pi}_k \hat{\pi}_k - \hat{\pi}_k(x) \Big(\del[S_E]{\phi_k}\Big) 
 \end{equation}
and will be called the stochastic Hamiltonian. The quantum theory defined by the Hamiltonian \(\mathcal{H}\) is a field theory in five dimensions, with the fictitious time \(\tau\) playing the role of time in fifth dimension. The equal time Green functions of this five dimensional field theory, in the equilibrium limit, will conicide, at least in perturbation theory, with the four dimensional CQFT.
The action corresponding to the five dimensional field theory is then seen to be
\begin{equation}
 \Lambda = \int \mathcal{L} dx d\tau, \qquad \text{where } \mathcal{L} = \left( \pi_k \dd[\phi_k]{\tau} -\mathcal{H} \right).
\end{equation}
\(\Lambda\), in the above equation, will be called stochastic action and \(\mathscr{L}\) will be called the stochastic Lagrangian. The operator formalism makes all the tools and techniques of conventional quantum field theory available for a study of the five dimensional field theory defined by the stochastic Lagrangian.

The formulation of SQM presented in this section is not the most general one.
One can  set  up a Langevin equation with a kernel and with  corresponding modification  
in the properties of the white noise. It is known that a kernel becomes necessary for fermionic models if the theory is to be renormalizable by power counting.

If the SQM of field theoretic models in four dimensions, is to be used as a framework of quantization,  the questions related to the ultra violet  divergences and renormalization must  be studied in the five dimensional theory itself.
This has been carried out extensively and we refer the reader to the papers published in Progress of Theoretical Physics Supplement in \cite{Sand}.  and references therein.  

%----------------------------------------------------------------------
% \input{SQM-Sec4-Mar23}

\section{Anomalous Axial Vector \(U(1)\) Gauge Model}
In this section we shall briefly recall the results of SQM study of a  model  of an axial vector \(U(1)\) gauge field coupled to a fermion. For SQM of fermionic theories with a kernel see, for example, \cite{Sand}.

%---------------------------------------------------------------------
The SQM is formulated in terms of the Euclidean action.
The {\it Euclidean action} for the
\(U(1)\) axial vector gauge theory will be taken to be  
\begin{eqnarray}
   S_E &=& \int d^4 x \mathscr{L} \\
   \mathscr{L} &=&  \bar{\psi}(-i\gamma_\mu D_\mu + m)
\psi - \frac{1}{4} F_{\mu\nu}F_{\mu\nu} + \frac{M^2}{2}A_\mu A_\mu  -
\frac{1}{2\alpha}(\partial\cdot A)^2
\end{eqnarray}
where
\begin{eqnarray}
 D_\mu = \partial_\mu -ig\gamma^5 A_\mu,\qquad F_{\mu\nu} = \partial_\mu A_\nu-
\partial_\nu A_\mu,
\end{eqnarray}
We will take the mass of the fermion to  be zero, \(m=0\). 
%===========================================================
\subsubsection*{Operation formalism}
Letting \(\pi_\mu,
\bar{\omega},\omega\) to denote the stochastic momenta corresponding to
the gauge field \(A_\mu\) and the fermionic fields \(\psi, \bar{\psi}\),
the stochastic action \(\Lambda\) of the five dimensional field theory takes
the following form:
\begin{equation}
 \Lambda = \int dx d\tau\left( \pi_\mu \frac{\partial A_\mu}{\partial
t}  + \frac{\partial\bar{\psi}}{\partial t}\omega  + \bar{\omega}
\frac{\partial\psi}{\partial t} - {\mathcal H}
 \right) \label{EQ07},
\end{equation}
where
\begin{eqnarray}
{\mathcal H}&=& \left[ \gamma^{-1} \pi_\mu\pi_\mu + 2\bar{\omega}K \omega
-\bar{\omega}\tilde{K}\frac{\delta S_E}{\delta \bar{\psi} } + \frac{\delta
S_E}{\delta \psi}\tilde{K} \omega -
\gamma^{-1}
\pi_\mu\frac{\delta S_E}{\delta A_\mu}
\right] \label{EQ08}.
\end{eqnarray}

Here \(K(x,t)\) is a suitable kernel that  needs to be used for the fermions.
For our present discussion an explicit expression for the kernel for fermions
is not required. See \cite{MPLA19} for more details.

A term
\(\frac{1}{2\alpha}(\partial_\mu A_\mu)^2\) has been included so that the
degree of divergence remains bounded.

\subsubsection*{Ultra violet divergences and counter terms}
A study of ultra violet divergences of the stochastic field theory defined by \eqref{EQ07} reveals that the finiteness of the theory requires addition of a new counter term 
\begin{equation}
\mathscr{L}_{\pi-A-A} = f\epsilon_{\mu\nu\alpha\beta} \pi_\mu (\partial_\nu A_\alpha) A_\beta
\end{equation}
Such a term is not present in the starting stochastic Hamiltonian \(\mathcal{H}\) of Eq \eqref{EQ08}. 

In previous paper \cite{MPLA19}, we have  obtained a Ward identity which
ensures that unphysical degree of freedom, the longitudinal  component of the
axial vector  field, decouples if  a suitable choice of the coupling constant \(f\) is made.

\subsubsection*{The counter term and Langevin formulation}

In the Langevin formulation  the effect of  addition of 
\(\pi-A-A\) counter term means modifying the Langevin equation for the 
gauge boson to \cite{akk20}
\begin{eqnarray}
  \frac{\partial A_\mu(x,t)}{\partial t}
  &=& - \gamma^{-1} \left(\frac{\delta S_E}{\delta A_\mu} +
\eta_\mu(x,t) - f\epsilon_{\mu\nu\lambda\sigma} A_\nu 
\partial_\lambda A_\sigma\right)  + \eta_\mu \label{EQ02}
\end{eqnarray}
and the Langevin equations for the fermions remain unchanged.
\begin{eqnarray}
\frac{\partial \psi(x,t)}{\partial t}
 &=& - \int dx^\prime K(x,x^\prime) \frac{\delta S_E}{\delta
\bar{\psi}(x^\prime)} + \theta(x,t)\label{EQ03A} \\
\frac{\partial \bar{\psi}(x,t)}{\partial t}
 &=&  \int dx^\prime  \frac{\delta S_E}{\delta {\psi}(x^\prime)} K(x,x^\prime)
+ \bar{\theta}(x,t)\label{EQ04A}.
\end{eqnarray}
The above Langevin form of SQM of   the \(U((1)\) axial vector gauge  model suggests   
that the  classical equation of motion be modified from 
\begin{equation}
 \frac{\partial S}{\partial A_\mu(x)} =0 
\end{equation}
to 
\begin{equation}
 \frac{\partial S}{\partial A_\mu(x)}  - f\epsilon_{\mu\nu\lambda\sigma} A_\nu 
\partial_\lambda A_\sigma =0
\end{equation}

These  EOM motion cannot be derived from a variational principle in four 
dimensions. That the equations of motion do not follow from an underlying  
Lagrangian is a major break from existing  quantum field 
theory models. 
%----------------------------------------------------------------------
We ask what does this \(\Pi-A-A\)  term correspond to in the equilibrium limit?  Can we go back to four dimensional  CQFT version of the model? If yes, what new features appear in the CQFT model?  The answers to some of these questions are as follows.

The appearance of \(\Pi-A-A\) term from one loop corrections means that the stochastic supersymmetry is broken and the correspondence  with the CQFT  model is lost.

The axial vector model of the previous paper \cite{MPLA19} holds promise of opening up the doors for other similar models which may be useful in going beyond standard model. Therefore, we ask  "How can computations of Green function and other quantities of interest be done without use of the operator formalism?" It will be useful  to study 
quantization of the axial vector gauge field and other similar models  directly in four dimensions. This will bypass the need to work with five dimensional stochastic quantum field and will lead to enormous  computational simplifications. It is not our intention here to develop, from scratch, a scheme of quantization based on the equations of motion. 
We continue from the work of the previous paper \cite{MPLA19} and suggest a scheme based on the Fokker Planck equation.

\section{Generating functional in four dimensions}
Starting from the operator formalism,  we wish to derive the equation for the generating functional of the theory in four dimensions.  It is expected that using resulting equations for the generating function will simplify the computations.  This is due to the fact that  one need not compute the correlation functions in the five dimensional stochastic field theory and  take the equilibrium limit.

With our objective as outlined above, we start with the Fokker Planck equation formalism of SQM. We wish to  derive an equation for the generating function 
\(Z[J]\). Computing required Green functions  directly in four dimensions, in a manner close to CQFT, will have the advantage of being much simpler than the corresponding computation of the equilibrium limit of correlation functions of the five dimensional theory described by the stochastic action. It is suggested that this objective can be achieved by starting with Fokker Planck equation and deriving 
an equation for the generating functional \(Z[J]\) of Green function of interest. 

% }\end{FileText}
% \end{document}

We write a general form of Fokker Planck Hamiltonian as
\begin{equation}\label{EQ30A}
 H_{FP} = \sum_k \frac{1}{\gamma_k} \int dx \del{\phi_k(x)}\left\{\del{\phi_k(x)}+\del[S_E]{\phi_k(x)} + E_k(\phi(x))\right\} 
\end{equation}
where \(E_k(\phi)\) represents extra terms corresponding to  counter terms that may arise as counter terms in higher orders.

 Let  \(P[\phi,\infty]\) denote the equilibrium limit of the Fokker Planck distribution function  \(P[\phi,\tau]\) as \(\tau\to \infty\). Then the equation satisfied by the equilibrium distribution function is
 \begin{equation}\label{EQ300}
  H_{FP} P[\phi,\infty] =0.
\end{equation}
Next, multiply \eqref{EQ300} by \(e^{i\int J\phi\,dx}\) and integrate over all field configurations. This gives
\begin{equation}
 \int \big( \Pi_m \mathcal{D} \phi_m\big) ~ e^{i\int J_n(y)\phi_n(y) dy}~
 H_{FP} P[\phi, \infty] = 0.\label{EQ32}
 \end{equation}

 By means of integration by parts, the functional derivatives w.r.t. the fields, \(\del{\phi_x(x)}\), can be transferred to the exponential, \(e^{i\int J_k(x)\phi_k(x) dy}\).  This gives
%  Doing this for the first functional derivative sitting, outside the square brackets in the Fokker Planck Hamiltonian \eqref{EQ30A}, gives  
 \begin{equation}\label{EQ30B}
 \int \big(\Pi_m \mathcal{D} \phi_m\big)  e^{i\int J_n(y)\phi_n(y)}~  \left[ \sum_k \frac{1}{\gamma_k}   \int dx J_k(x)~
 \left\{-iJ_k(x)+\del[S_E]{\phi_k(x)} + E_k(\phi(x))\right\} ~\right] H_{FP}  =0 .
\end{equation}
 
 Next, using integration by parts, the fields appearing inside the curly brackets in \eqref{EQ30B}, can be written as 
functional derivatives \(\frac{1}{i} \del{J_k}\) on the exponential 
\( e^{i\int J_k(y)\phi_k(y) dy}\). These  functional derivatives, w.r.t. \(J\) can be pulled out of the functional integration. The \eqref{EQ32} then gets transformed into a functional differential equation 
\begin{equation}\label{EQ30}
 \left[ \sum_k \frac{1}{\gamma_k}~\int dx
 \left\{-iJ_k(x)+\del[S_E]{\phi_k(x)} + E_k(\phi(x))\right\}_{\phi\to  \delta_J}  J_k(x)~ \right] Z[J] =0 ,
\end{equation}
where the notation \(\phi\to  \delta_J\) is a short hand for the replacement 
\(\phi(x)\to \frac{1}{i} \del{J(x)}\), and \(Z[J]\) is the generating function for the Green functions of the theory 
\begin{equation}
 Z[J] \stackrel{\text{def}}{=} 
 \int \Pi_m \big(\mathcal{D} \phi_m\big) ~ e^{i\int J_n(y)\phi_n(y) dy}~
P[\phi, \infty].\label{EQ312}
 \end{equation}

It is suggested that  \eqref{EQ30} for the generating function should be the starting point of perturbative computations of interest. In most practical applications, it would be sufficient to obtain the Green functions up to lowest one or two orders in the term \(E(\phi)\), representing extra counter terms which cannot be absorbed into redefinition of the Euclidean action.   

\section{Concluding Remarks}
The approach to the anomalous  axial vector gauge model, suggested here is a promising one. It is equivalent to the operator formalism and permits computations directly in four dimensions. This method has great promise  for study of similar models potentially useful for going beyond the standard model.   

What are possible consequences of this result for going beyond the standard model remains to be investigated. Two problems,  requiring immediate attention, are the problem of neutrino masses and the CP violation. It has been argued that the anomalous axial vector theory may provide a mechanism for neutrino mass through  radiative corrections.
The extra counter terms, that arise in one loop, may give rise to CP violation in a suitably constructed model.

Taking the investigation to  a logical conclusion, in a realistic model, appears to be too complicated. There is an urgent need to have techniques to achieve as much simplification as possible.  
The proposals here may well provide an important framework to study a class of new models beyond the standard model of elementary particles.

\paragraph*{Acknowledgment}
I thank Subhash Chaturvedi and E. Harikumar for going through an earlier version of the manuscript and suggesting several improvements.  
%==========================================================


\begin{thebibliography}{123456}
\bibitem{MPLA19}
A.K. Kapoor, Modern Physics Letters {\bf A 34}, No. 22, 1950176 (2019)\\
https://doi.org/10.1142/S0217732319501761
\bibitem{akk20} A. K. Kapoor, "SQM of Anomalous \(U(1)\) Axial Vector Gauge Theory"
unpublished (2020).
\bibitem{akk15}
A.K. Kapoor, "Radiative Corrections as Origin of Tiny Fermion Masses",
arXiv:1502.07.07637v1 [hep-th] (2015) and references therein.
\bibitem {SQM1} G. Parisi and Y. Wu Sci. Sinica {\bf 24}  (1981) 483
\bibitem {SQM2}  M. Namiki, `` Stochastic Quantization'' Springer Verlag, New
York (1992).
\bibitem {SQM3}  S.  Chaturvedi, A. K. Kapoor and V. Srinivasan, ``Stochastic
Quantization Scheme of Parisi and Wu'', Bibliopolis, Napoli (1990).
\bibitem {SQM4} Daamgard and H\"{u}ffel,``Stochastic Quantization'', World
Scientific, Singapore (1988).
\bibitem{Sand}
R. Sandhya, S. Chaturvedi, A. K. Kapoor and V. Srinivasan, Progress of
Theoretical Physics Supplement Number 111 (1993) 237.\\
This issue of  Progress of Theoretical Physics Supplement Number {\bf 111}  is devoted to the stochastic quantization method. An interested reader will find many useful reviews and articles.
\end{thebibliography}
\end{document}